\documentclass{article}
\usepackage{spconf,amsmath,graphicx}

\usepackage{booktabs}       
\usepackage{amsfonts}       
\usepackage{nicefrac}       
\usepackage{microtype}      
\usepackage{multirow}
\usepackage{lipsum}
\usepackage{url}
\usepackage{pifont}
\usepackage{fancyhdr}       
\usepackage{graphicx}       
\graphicspath{{media/}}     
\usepackage{soul, color, xcolor}
\usepackage{makecell}

\usepackage[a4paper, total={7.2in, 9.2in}]{geometry}

\begin{document}\sloppy
\title{Multi-Dimension-Embedding-Aware Modality Fusion Transformer for Psychiatric Disorder Clasification}
%

\name{\makecell*[c]{Guoxin Wang$^{1}$, Xuyang Cao$^{2}$, Shan An$^{2}$, Fengmei Fan$^{3}$, \\ 
Chao Zhang$^{2}$, Jinsong Wang$^{2}$, Feng Yu$^{1,\dag}$, Zhiren Wang$^{3}$
\thanks{$^{\dag}$Corresponding authors}}}
\address{$^{1}$College of Biomedical Engineering \& Instrument Science, Zhejiang University, Hangzhou, China, \\
$^{2}$JD Health International Inc., Beijing, China, \\
$^{3}$Beijing Huilongguan Hospital, Peking University Huilongguan Clinical Medical School, Beijing, China.}
\maketitle

%
%
%

%
\begin{abstract}
Deep learning approaches, together with neuroimaging techniques, play an important role in psychiatric disorders classification.
Previous studies on psychiatric disorders diagnosis mainly focus on using functional connectivity matrices of resting-state functional magnetic resonance imaging (rs-fMRI) as input, which still needs to fully utilize the rich temporal information of the time series of rs-fMRI data.
In this work, we proposed a multi-dimension-embedding-aware modality fusion transformer (MFFormer) for schizophrenia and bipolar disorder classification using rs-fMRI and T1 weighted structural MRI (T1w sMRI). Concretely, to fully utilize the temporal information of rs-fMRI and spatial information of sMRI, we constructed a deep learning architecture that takes as input 2D time series of rs-fMRI and 3D volumes T1w. Furthermore, to promote intra-modality attention and information fusion across different modalities, a fusion transformer module (FTM) is designed through extensive self-attention of hybrid feature maps of multi-modality. In addition, a dimension-up and dimension-down strategy is suggested to properly align feature maps of multi-dimensional from different modalities. Experimental results on our private and public OpenfMRI datasets show that our proposed MFFormer performs better than that using a single modality or multi-modality MRI on schizophrenia and bipolar disorder diagnosis.
\end{abstract}
\section{Introduction}
\label{sec:intro}
Accurate diagnosis of psychiatric disorders plays a crucial role in providing an opportunity for appropriate treatments and potentially evaluating the effectiveness of treatments \cite{takashi_structured_2018}. Currently, psychiatric disorders are diagnosed mainly relying on behavioral symptoms according to the Diagnostic and Statistical Manual of Mental Illnesses (DSM-5) \cite{dsm-5}. Nevertheless, machine learning based methods with brain imaging techniques have great potential in providing more stable and precise diagnoses and even finding biological or physiological biomarkers to reason about certain disorders \cite{li2022data,yan_2022_mapping}.

\begin{figure*}[!ht]
    \centering
    \includegraphics[width=0.95\textwidth]{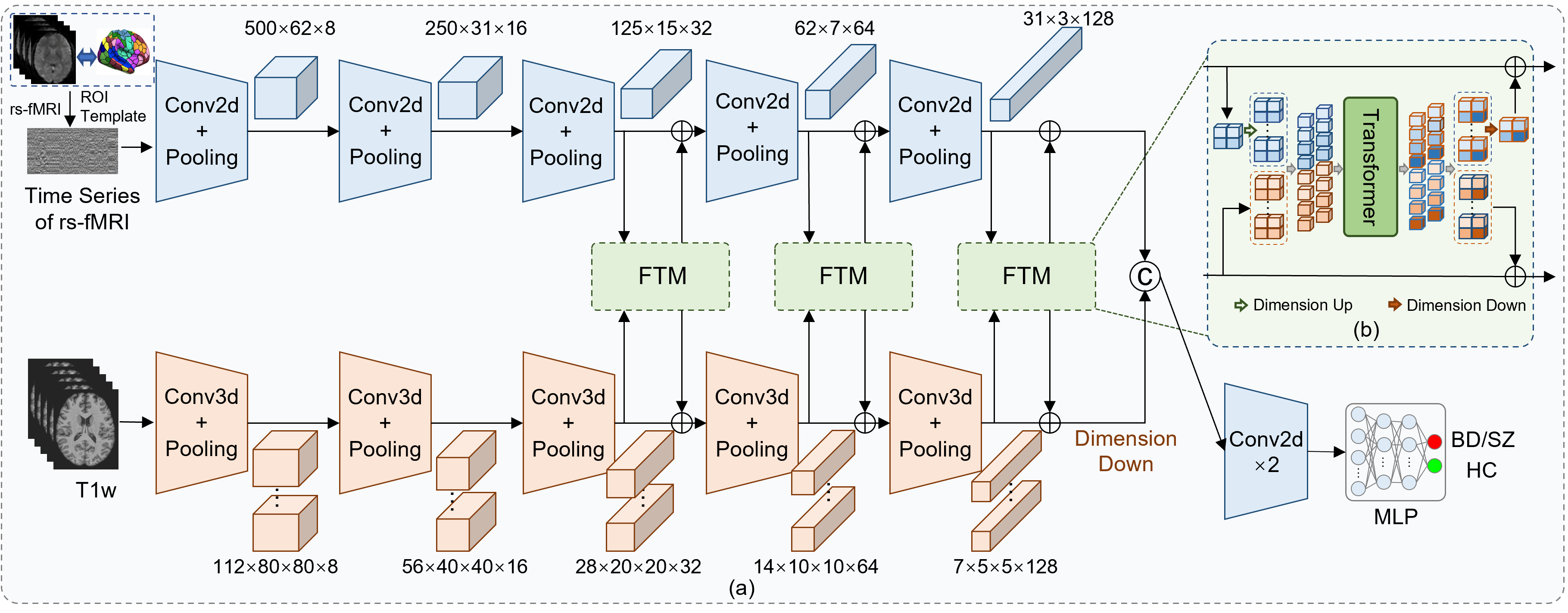}
    \caption{An overview of the proposed MFFormer.}
    \label{fig:model_arch}
\end{figure*}

A lot of current machine learning based approaches distinguish patients with psychiatric disorders from healthy controls utilizing a single modality \cite{rivera_2022_diagnosis, zhu2022application}.
However, as multi-modality data is able to provide richer information, deep learning methods with multi-modality fusion is a promising direction to improve the model performance in a variety of medical image analysis tasks \cite{mmFormer_2022,zheng2022multi,sun2022two}. 
Although recent multi-modal study for Alzheimer’s disease diagnosis demonstrated that a single-modality network using FDG-PET performs better than MRI and does not show improvement when combined \cite{marla_isapet_2022}, the fusion way they used was simply concatenating features from different modalities in the early/middle or late stage. We believe that the fusion strategy used in \cite{marla_isapet_2022} is not able to fully explore the complex dependencies between modalities. Consequently, a more effective modality fusion mechanism should be carefully designed to better model the complex non-linear dependencies across modalities.

In this paper, we proposed a multi-dimensional-embedding-aware modality fusion transformer (MFFormer) for diagnosing schizophrenia (SZ) and bipolar disorder (BD) using resting state fMRI (rs-fMRI) and T1 weighted (T1w) sMRI. Concretely, to make full use of the temporal information of rs-fMRI and spatial information of sMRI, we constructed a deep learning architecture that takes as input 2D time series of rs-fMRI and 3D volumes of T1w sMRI. In addition, a fusion transformer module (FTM) is designed to encourage the network fully explore the non-linear dependencies between modalities.
Furthermore, to ensure feature fusion across different dimensions, we suggested a dimension-up and dimension-down operation before the 3D feature maps of time series in and out of the FTM. Thus, the multi-dimensional can be appropriately aligned during modality fusion. Extensive experiments on our private and public OpenfMRI datasets demonstrated that our proposed MFFormer significantly outperformed existing approaches and improved compared to a single modality.

\section{Methods}
\label{sec:format}
The overall structure of the proposed MFFormer is illustrated in Fig. \ref{fig:model_arch}. Two branches of encoders are constructed individually for the 2D time series of rs-fMRI and the 3D T1w volume. The non-linear dependencies between different modalities are extensively explored between two branches using the designed FTM. After five convolution layers, the 4D embeddings of T1w were down-sampled to 3D, and then feature maps from each branch are aggregated with concatenation. Finally, the aggregated features are fed into two convolution layers and a multi-layer perceptron (MLP), and output the final result.

\subsection{Feature Extraction Layer}
As seen in Fig. \ref{fig:model_arch} (a), the structure of encoders of both input modalities is the same except for the dimensions of operations, including convolution, batch normalization, and pooling. The notion "Conv2d + pool" indicates a 2D convolution followed by batch normalization and a LeakyReLU, then a max pooling with a stride of 2 is used to half the spatial resolution of feature maps for a higher level of representations. The same procedure is defined for "Conv3d + pool" while all the operations are changed to 3D. Each encoder's channels of feature maps are doubled after passing through the convolutional layer. 

Note that for the T1w, the preprocessed data are input directly into the encoder. For the rs-fMRI, the 4D preprocessed data is first parcellated into 1000 regions of interest (ROIs) using the Schaefer 2018 parcellation template \cite{schaefer_local_2018}, then the yield 2D time series image is used as input of the 2D encoder. In this way, more temporal information could be explored than the traditional FNC matrix based input. The dimension of the 2D time series is $t\times 1000$, where $t$ is the number of frames in each rs-fMRI. In this paper, $t$ was set to 124 after preprocessing.

\subsection{Fusion Transformer Module}\label{sec:ftm}
The key idea of the proposed FTM is to explore the complex non-linear dependencies between different modalities using the self-attention mechanism of transformer \cite{ashish_attention_2017}. In this way, features within modality would be augmented, and information from different modalities can be better aggregated to promote the final classification results.
Fig. \ref{fig:model_arch} (b) shows details of the proposed FTM. First, the FTM takes as input multi-dimension feature maps from two modalities. 
Second, the aligned feature maps are fused and then fed into the transformer for better attention within intra-modality and inter-modality. Finally, the output of the transformer is further reshaped to the original dimensions of feature maps and then aggregated back to each modality branch.

Formally, let the feature maps of rs-fMRI be $F_f \in \mathbb{R}^{W_f\times H_f\times C_f}$, and the feature maps of T1w be $F_s \in \mathbb{R}^{W_s\times H_s\times D_s \times C_s}$, respectively. Generally, there are two optional fusion approaches before feature maps from different dimensions can be aggregated: 1) up-sampling 3D feature maps to 4D; and 2) down-sampling 4D feature maps to 3D. The dimension-up strategy utilizes the first approach to avoid the loss of information. Concretely, the dimension of feature maps of rs-fMRI is up-sampled by repeating elements $D$ times, yielding a new feature map of $F_f \in \mathbb{R}^{W_f\times H_f\times D\times C_f}$. We further conduct average pooling to rearrange the two feature maps into the same shape of $\mathbb{R}^{W\times D\times H\times C}$, then flatten and stacked the feature maps together, yielding a tensor of $\tilde{F} \in \mathbb{R}^{2WHD\times C}$.

The stacked feature maps are fed as input to a transformer for extensive attention. Following existing methods on vision transformers \cite{kimberly_hybrid_2022}, position embedding was employed in this paper to enable the network to construct spatial dependence between different tokens. The attention mechanism used in this paper can be formulated as follows:
\begin{equation}
    Q = \tilde{F}W_q, K=\tilde{F}W_k, V=\tilde{F}W_v \\
\end{equation}

\begin{equation}
    \tilde{F}_{out} = \tilde{F} + SoftMax(\frac{QK^T}{\sqrt{d}}) V
\end{equation}
where $Q, K, V \in \mathbb{R}^{C\times d}$ are the weight matrices, $d=128$ are the dimension of $Q, K, V$. The dimension of output $\tilde{F}_{out}$ is $2WHD\times C$.

The output of the transformer is divided into two tensors with dimensions of $WHD\times C$, and then restored to the original shape of each input tensor through interpolation. The restored feature maps are finally aggregated back to their encoder branches using element summation. Note that a dimension-down strategy is conducted by averaging features of rs-fMRI along the $D$ axis before aggregating back, as illustrated in Fig. \ref{fig:model_arch} (b).

\subsection{Classification Layer}
In the classification layer, to properly fuse information from two modalities, feature maps of T1w were first averaged over $D_s$ dimension, yielding a dimension of $\mathbb{R}^{C_s\times W_s\times H_s}$. Then, an adaptive average pooling was used to adjust the size of T1w feature maps to be the same as that of rs-fMRI. After concatenating, the fused features can be represented as $F_{fused} \in \mathbb{R}^{2C_f\times W_f\times H_f}$. The fused features were first input into two convolution layers for better non-linear representations, then flattened and fed into a three-layer MLP for final output. In the classification layer, two settings were carefully designed to prevent overfitting on small datasets. First, the 4D feature maps of T1w were down-sampled to fit the 3D feature maps of rs-fMRI, which is not the same as that in the proposed FTM. Second, a smaller number of neurons was set (20 in this paper) in the two hidden layers of MLP. Finally, the cross-entropy loss function was employed for network training.

\section{Experiments and Results}

\subsection{Data Acquisition and Preprocessing}

Two datasets were utilized in this paper, including a private BD dataset and the public OpenfMRI dataset with the accession number of ds000030 (\url{https://openfmri.org/dataset/ds000030/}).

\textbf{BD dataset.} Our private BD dataset was collected from Psychiatric Hospital of Zhumadian from 2014 to 2018, with a total number of 91 subjects (52 BD patients and 39 HC). Each subject contains two-modality MRI data (T1w anatomical MPRAGE and rs-fMRI). The T1w image has a volume size of $256\times 256\times 188$, and the volume size of rs-fMRI image is $64\times 64\times 33$ with 210 frames.

\textbf{OpenfMRI dataset.} The publicly available OpenfMRI dataset with accession number ds000030 (revision version 1.05) was obtained, including 221 subjects (122 HC, 50 SZ subjects, and 49 BD patients). T1w anatomical MPRAGE image and rs-fMRI were used the same as our private BD dataset.

Both datasets were preprocessed using the fMRIPrep toolkit \cite{esteban2019fmriprep}. For the T1w data, skull stripping was first performed to extract brain regions from the full image, then spatial normalization was performed by registering images to the MNI 152 template.
The preprocessing of rs-fMRI includes skull stripping, head motion correction, susceptibility distortion correction, and spatial normalization.

\subsection{Evaluation Metrics}
Following existing methods for the psychiatric disorder diagnosis, four evaluation metrics were utilized to compare different classification models, including the F1 score, sensitivity (SEN), specificity (SPEC), and balanced accuracy (BACC). 

\begin{table*}[ht]
    \caption{classification results on the public Open-fMRI dataset. $\dagger$ indicates that results originate from the sw-DGM.}
    \footnotesize
    \centering
    \begin{tabular}{ccccccccc}
        \toprule
        \multirow{2}{*}{Method} & \multicolumn{4}{c}{Schizophrenia} & \multicolumn{4}{c}{Bipolar Disorder}\\
        \cmidrule(lr){2-5} \cmidrule(lr){6-9}
         & BACC & F1 & SEN & SPEC & BACC & F1 & SEN & SPEC\\
        \midrule
        PCC+SVM \cite{WeizhengYan2019DiscriminatingSU} & 0.595 & 0.361 & 0.342 & 0.848 & 0.555 & 0.403 & 0.422 & 0.689 \\
        PCC+MLP \cite{WeizhengYan2019DiscriminatingSU} & 0.633 & 0.440 & 0.372 & \textbf{0.894} & 0.605 & 0.323 & 0.239 & \textbf{0.963} \\
        LSTM$^{\dagger}$ \cite{nicha_identifying_2017} & 0.661 & - & 0.854 & 0.467 & 0.571 & - & 0.802 & 0.340 \\
        DGM$^{\dagger}$ \cite{takashi_deep_2017} & 0.722 & - & \textbf{0.920} & 0.524 & 0.619 & - & 0.650 & 0.587 \\
        sw-DGM$^{\dagger}$ \cite{takashi_structured_2018} & \textbf{0.767} & - & 0.812 & 0.722 & 0.622 & - & \textbf{0.844} & 0.401 \\
        Late Fusion \cite{marla_isapet_2022} & 0.756 & 0.635 & 0.686 & 0.826 & 0.657 & 0.496 & 0.767 &  0.548 \\ 
        MFFormer (ours) & 0.763 & \textbf{0.655} & 0.706 & 0.805 &  \textbf{0.782} & \textbf{0.673} & 0.773 & 0.792 \\
        \bottomrule
    \end{tabular}
    \label{tab:quantitative_result_public}
\end{table*}

\begin{table}
    \caption{Classification results on our private BD dataset.}
    \centering
    \begin{tabular}{ccccc}
        \toprule
        Method & BACC & F1 & SEN & SPEC \\
        \midrule
        PCC+SVM \cite{WeizhengYan2019DiscriminatingSU} & 0.570 & 0.595 & 0.621 & 0.519 \\
        PCC+MLP \cite{WeizhengYan2019DiscriminatingSU} & 0.595 & 0.547 & 0.545 & 0.646 \\
        Late Fusion \cite{marla_isapet_2022} & 0.686 & 0.542 & \textbf{0.774} & 0.597 \\
        MFFormer (ours) & \textbf{0.768} & \textbf{0.755} & 0.684 & \textbf{0.851}\\
        \bottomrule
    \end{tabular}
    \label{tab:quantitative_result_private}
\end{table}

\subsection{Implementation Details}
All the models in this paper were trained and tested using the Pytorch architecture on one Tesla P40. The training batch size was set to 5. The AdamW optimizer was used with a weight decay of 1e-8 and a momentum of 0.99. The learning rate was set to 5e-4, and a linear warm-up cosine learning rate scheduler was employed. The warm-up epochs of the scheduler were set to 200, and the total number of epochs was set to 500. It is worth noting that an over-sampling strategy was adopted during training, as a severe class imbalance problem exists in both datasets used in this paper. Specifically, a weighted sampler was utilized to keep positive and negative samples approximately the same in each training batch.

\subsection{Comparison with existing Methods}
To verify the effectiveness of our proposed method, we compare the classification results of our proposed MFFormer with several existing state-of-the-art methods. For both datasets, we implement the classic method that calculates the FNC matrix of rs-fMRI as input features \cite{WeizhengYan2019DiscriminatingSU}, then train the classification models using the support vector machine (SVM) or MLP. We also implement the late fusion method that performs the best among three multi-modality fusion strategies in reference \cite{marla_isapet_2022}. Additionally, for the public OpenfMRI dataset, we report more results using the single rs-fMRI data (LSTM \cite{nicha_identifying_2017}, DGM \cite{takashi_deep_2017}, sw-DGM \cite{takashi_structured_2018}). For a fair comparison, all the networks are trained and tested using the same parameters except the network architectures. Five-fold cross-validation is used with the same train-test split for all the approaches.

\subsubsection{Comparison on BD dataset}
Table \ref{tab:quantitative_result_private} shows different classification results on our private BD dataset. Our proposed MFFormer significantly outperforms existing methods in single-modality and multi-modality fusion schemes, with a BACC of 0.768 and an F1 score of 0.755. Besides, the proposed MFFormer achieves a better balance between sensitivity (0.684) and specificity (0.851) than all other methods. It is worth noting that the encoder of T1w and rs-fMRI are the same in the Late Fusion \cite{marla_isapet_2022} and MFFormer, while the MFFormer achieves a better BACC than the Late Fusion by a large margin (8.2\%).

\subsubsection{Comparison on OpenfMRI dataset} Table \ref{tab:quantitative_result_public} reports classification results of different methods on the public Open-fMRI dataset. Following existing approaches, we conduct SZ and BD diagnosis separately using two individual classification models. In the SZ vs. HC classification, our proposed method achieves the second-best performance in terms of BACC (0.763), and achieves the best F1 score of 0.655 among all the methods. In the BD vs HC classification, the Late Fusion method achieves the best BACC of 0.657 and F1 score of 0.496 among all the compared methods. Compared with the Late Fusion, our proposed MFFormer achieves a better BACC and F1 score, with improvements of 12.5\% and 17.7\%. 


\subsection{Ablation Study}
We conducted ablation studies on the OpenfMRI dataset (BD) to evaluate the proposed FTM's effectiveness and several modality fusion schemes, as seen in Table \ref{tab:ablation_study}. Baseline 1 and baseline 2 are models that only utilize T1w or rs-fMRI without modality fusion, respectively. We further compared different modality-fusion strategies (described in subsection \ref{sec:ftm}) before features from different dimensions are aggregated: 1) up-sample 3D feature maps to 4D; and 2) down-sample 4D feature maps to 3D. Consequently, the "3D 1-way" denotes the first modality-fusion strategy and only utilizes one FTM in the last layer of T1w and rs-fMRI encoders. The "3D 3-way" denotes the first modality-fusion strategy and utilized 3 FTMs in the last three layers of T1w and rs-fMRI encoders. The "4D 1-way" and "4D 3-way" represent the second modality-fusion strategy with 1 or 3 FTMs, respectively.

\begin{table}
\renewcommand\tabcolsep{3.8pt}
    \caption{Ablation study on the public OpenfMRI dataset (BD).}
    \resizebox{\columnwidth}{!}{
    \centering
    \begin{tabular}{ccccccccc}
        \toprule
        Model & T1w & rs-fMRI & fusion method & BACC & F1 & SEN & SPEC \\
        \midrule
        baseline 1 & \ding{51} & \ding{55} & \ding{55} & 0.542 & 0.383 & 0.500 & 0.584 \\
        baseline 2 & \ding{55} & \ding{51} & \ding{55} & 0.701 & 0.580 & 0.730 & 0.673 \\
        MFFormer & \ding{51} & \ding{51} & 3D 1-way & 0.706 & 0.552 & 0.541 & \textbf{0.871} \\
        MFFormer & \ding{51} & \ding{51} & 3D 3-way & 0.716 & 0.583 & 0.768 & 0.663 \\
        MFFormer & \ding{51} & \ding{51} & 4D 1-way & 0.763 & 0.648 & 0.705 & 0.820 \\
        MFFormer & \ding{51} & \ding{51} & 4D 3-way & \textbf{0.782} & \textbf{0.673} & \textbf{0.773} & 0.792 \\
        \bottomrule
    \end{tabular}
    }
    \label{tab:ablation_study}
\end{table}

It can be seen in Table \ref{tab:ablation_study} that all the models with modality-fusion perform better than that with only a single modality, indicating the effectiveness of the modality-fusion mechanism. Compared with "3D x-way" fusion method, the "4D x-way" method achieves significantly better BACC and F1 scores. This result may be the loss of information due to down-sampling 4D features to 3D. Compared with "3D 1 way" and other methods, the modality fusion with the "3D 3-way" strategy achieves the best performance in terms of BACC (0.782) and F1 score (0.673), indicating more FTMs are useful in modality fusion.

\section{Conclusion}
This paper proposes a multi-dimensional embedding-aware modality fusion transformer, denoted as MFFormer, to diagnose schizophrenia and bipolar disorder. The proposed MFFormer takes rs-fMRI time series and T1w sMRI as inputs, and then the complex non-linear dependencies between modalities are explored using our designed FTM by aggregating feature maps from different dimensions. Extensive experimental results on our private BD dataset and a public OpenfMRI dataset demonstrate that our proposed MFFormer outperforms models with a single modality and existing methods for psychiatric disorder diagnosis. In the future, more effective encoders of fMRI and T1w based on transformer will be explored.


\bibliographystyle{IEEEbib}
\bibliography{strings,refs}

\end{document}